\documentclass[fleqn,10pt]{wlscirep}
\usepackage{braket}
\usepackage[onehalfspacing]{setspace}
\title{Highly indistinguishable and strongly entangled photons from symmetric GaAs quantum dots}

\author[1]{Daniel Huber*}
\author[1]{Marcus Reindl}
\author[1]{Yongheng Huo}
\author[1]{Huiying Huang}
\author[1]{Johannes S. Wildmann}
\author[2]{Oliver G. Schmidt}
\author[1]{Armando Rastelli*}
\author[1]{Rinaldo Trotta*}
 
\affil[1]{Institute of Semiconductor and Solid State Physics, Johannes Kepler University, Linz,
Altenbergerstraße 69, 4040, Austria}
\affil[2]{Institute for Integrative Nanosciences, IFW Dresden, Helmholtzstraße 20, 01069 Germany}

\affil[*]{corresponding author:daniel.huber@jku.at, armando.rastelli@jku.at, rinaldo.trotta@jku.at}

\begin{abstract}

The development of scalable sources of non-classical light is fundamental to unlock the technological potential of quantum photonics\cite{Kimble:Nat2008}. Among the systems under investigation\cite{PhysRevLett.75.4337,Hacker:Nat2016,Igor:Nat2016}, semiconductor quantum dots are currently emerging as near-optimal sources of indistinguishable single photons\cite{Senellart:NatPhoton2016,PhysRevLett.116.020401}. However, their performances as sources of entangled-photon pairs are in comparison still modest. Experiments on conventional Stranski-Krastanow InGaAs quantum dots have reported non-optimal levels of entanglement and indistinguishability of the emitted photons\cite{Michler:NaturePhoton2014,PhysRevLett.108.040503,Trotta:NanoLet2014,Dousse:Nat2010}. For applications such as entanglement teleportation and quantum repeaters, both criteria have to be met simultaneously\cite{Kimble:Nat2008,ZollerNat2001}. In this work, we show that this is possible focusing on a system that has received limited attention so far: GaAs quantum dots grown via droplet etching. Using a two-photon resonant excitation scheme\cite{Michler:NaturePhoton2014}, we demonstrate that these quantum dots can emit triggered polarization-entangled photons with high purity (g$^{(2)}(0)=0.002 \pm0.002$), high indistinguishability ($0.93 \pm 0.07$) and high entanglement fidelity ($0.94 \pm0.01$). Such unprecedented degree of entanglement, which in contrast to InGaAs can theoretically reach near-unity values, allows Bell's inequality ($2.64 \pm0.01$) to be violated without the aid of temporal or spectral filtering. Our results show that if quantum-dot entanglement resources are to be used for future quantum technologies, GaAs might be the system of choice.

\end{abstract}

\begin{document}

\flushbottom
\maketitle
%
%
\thispagestyle{empty}

\section*{Introduction}

Any prototype of quantum device that will be used in emerging quantum technologies has to fulfill a long list of requirements. The ideal source of quantum light, for example, should deliver single and entangled photons deterministically, with high purity, high efficiency, high indistinguishability, high degree of entanglement, and it should be compatible with current photonic integration technologies. Implementing such a source, however, is a task far from being easy\cite{LU:2014NatPhoton}. 

If we restrict the discussion to single photon sources\cite{Igor:Nat2016}, i.e., we disregard entanglement, semiconductor quantum dots (QDs) have recently demonstrated their capability to fulfill all the requirements of the "wish-list", thus carrying great promise for the implementation of photonic quantum networks\cite{Kimble:Nat2008}. This is eventually testified by the increasing interest of the community working with parametric down converters\cite{Loredo:arXiv2016,He:arXiv2016}, the workhorse sources of non-classical light that have been used to achieve a wealth of breakthroughs in quantum optics.

Besides single photons, QDs can also generate triggered polarization-entangled photons during the biexciton (XX)-exciton (X) radiative cascade\cite{PhysRevLett.84.2513}. Having in mind their performances as single photon sources, it would be natural to regard QDs as ideal entanglement resources.
The reality, unfortunately, is quite distant from this idea. The main reason is that the development of scalable sources of entangled photons demands for approaches that are much more advanced than those employed for single photons. First, the very possibility to use a QD to generate entangled photons poses severe constraints on its structural symmetry\cite{PhysRevLett.109.147401}. Second, the efficient extraction of both X and XX photons requires sophisticated photonic micro-cavities or broad-band nanowires\cite{Versteegh:NatCom2014,Dousse:Nat2010}. Third, two-photon resonant excitation schemes are needed to achieve truly on-demand generation of photon-pairs and to avoid some of the decoherence processes limiting the indistinguishability of the emitted photons\cite{Michler:NaturePhoton2014,PhysRevLett.110.135505}. Despite the different points have been recently addressed (although never simultaneously on the very same device), non-ideal levels of indistinguishability and degree of entanglement were reported so far\cite{Michler:NaturePhoton2014,PhysRevLett.108.040503}. In particular for the latter parameter, all the experiments have shown fidelities to the expected Bell state that rarely exceed 80$\%$ (without inefficient post-selection techniques)\cite{Kuroda:2013PRB,Versteegh:NatCom2014,Dousse:Nat2010,Trotta:NanoLet2014,PhysRevLett.108.040503}, even under resonant pumping\cite{Michler:NaturePhoton2014}. Assuming that the X states involved in the cascade are degenerate\cite{Trotta:NanoLet2014} and ultra-clean QD sample with reduced charge noise\cite{Kuhlmann:2013NatPhys} are studied, the degree of entanglement is mainly limited by (i) recapture processes\cite{Kuroda:2013PRB,Dousse:Nat2010} and (ii) random magnetic fields produced by the QD nuclei\cite{Vandersypen:NatMat2013,Burk:arxiv2015}. (i) arises when the intermediate X levels are re-excited to the XX level before decaying to the ground state. (ii) is instead related to the hyperfine interaction\cite{Atature:NatureCom2016}, which couples the spin of the electron and hole forming the exciton with the spins of the QD nuclei, effectively giving rise to a fluctuating magnetic field (Overhauser field) which depolarizes randomly the X states.
It is well known that these sources of entanglement degradation lower the maximum entanglement fidelity, and there are strategies to compensate for their deleterious effect, at least in principle.

Recapture processes can be suppressed by shortening the X decay time via the Purcell effect\cite{Dousse:Nat2010} or/and using resonant excitation\cite{Michler:NaturePhoton2014,PhysRevLett.110.135505}. Nuclear spin-polarization techniques\cite{Munsch:NatNano2014} can be used to polarize the nuclear ensemble with the drawback of inducing a finite effective magnetic field which splits the X states\cite{Bayer:PRB2014}. This splitting can be in turn compensated via additional external perturbations, such as strain, magnetic or electric field\cite{Trotta:2016NatCom,PhysRevLett.109.147401}. Even if theoretically feasible, this approach is not straightforward to implement experimentally, as the effect of a magnetic field on the X states depends strongly on its direction and can give rise to off-diagonal terms in the X Hamiltonian that cannot be easily compensated with external fixes. 

In this work, we look at this problem from a different perspective. Instead of using Stranski-Krastanow InGaAs QDs – the standard choice in QD photonics – we focus on highly symmetric GaAs QDs grown via the droplet etching method\cite{Huo:APL2013,APL1.3653804}. In this type of artificial atoms the effect of the Overhauser field is expected to be much weaker than in InGaAs QDs\cite{Vandersypen:NatMat2013}. This is not only due to the absence of Indium in the QD nuclei. GaAs/AlGaAs QDs are indeed free of intrinsic strain fields and composition gradients that have a relevant role on the hyperfine interactions\cite{Atature:NatureCom2016}. Moreover, the highly-symmetric shape of the investigated QDs minimizes the degree of light- and heavy-hole mixing, which also depolarizes the X states\cite{PhysRevB.94.121302}. The use of this novel QDs is not the only ingredient to push the degree of entanglement to its limits. Differently from previous attempts on droplet epitaxy GaAs QDs\cite{Kuroda:2013PRB}, we take advantage of two-photon resonant excitation schemes to coherently populate the XX state\cite{Michler:NaturePhoton2014}. On the one hand, this technique limits the effect of re-excitation. On the other hand, it allows us to demonstrate high visibilities of two-photon interference for consecutive photons, a key parameter that has never been investigated for the GaAs system.

\section*{Results and discussion}

\begin{figure}[htbp]
	\centering
		\includegraphics[width=90mm]{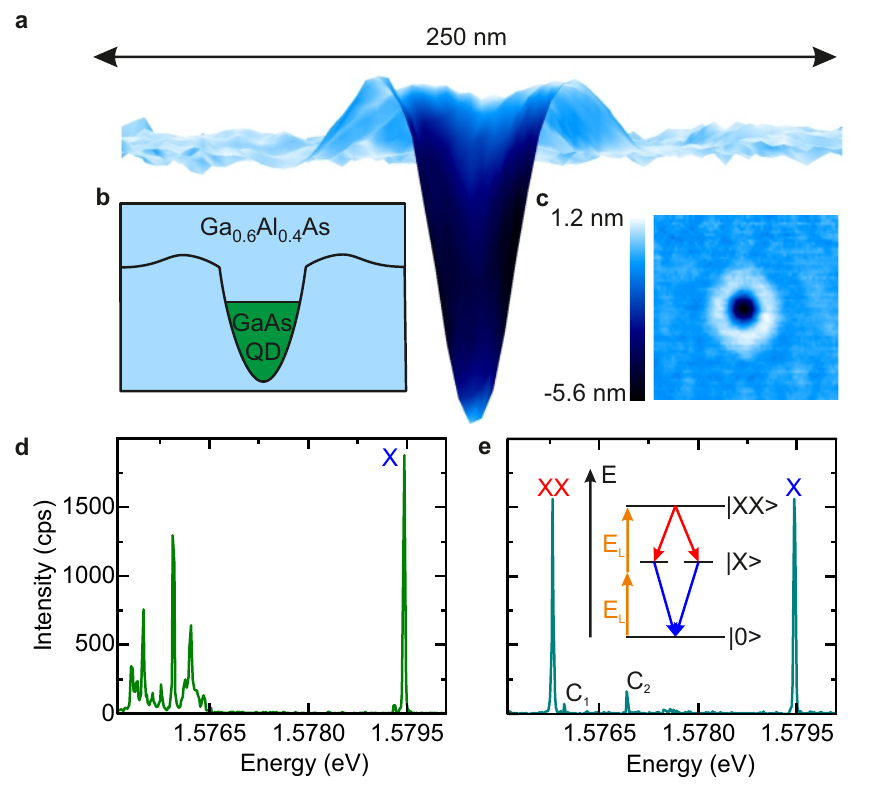}
	\caption{\textbf{Sample structure and spectral properties of the used QDs} (a) Cross-sectional 3D view of an Atomic Force Microscopy (AFM) image of a nanohole in an AlGaAs layer. A GaAs QD with a height of about 7 nm is obtained after its filling with GaAs and overgrowth with AlGaAs. (b) Sketch of the sample structure. (c) Top view of the AFM measurement. The highly symmetric shape is crucial to obtain the high degree of entanglement observed here. (d) $\mu-$PL spectrum of a representative QD under non-resonant excitation (laser photon energy $E_L = 2.54$ eV). The neutral exciton emission line is labeled as X. The XX is not visible in these excitation conditions. (e) Spectrum of the same QD as in (d) under resonant two photon excitation ($E_L=1.5775$ eV). The X and XX are visible and have very similar intensity. Two weaker charged states appear (C$_{1}$ and C$_{2}$).  Inset: For resonant two-photon excitation, the 9 ps laser pulses are tuned to half of the energy difference between XX and ground state $0$.}
	\label{fig:fig1}
\end{figure}

We begin with the characterization of our (GaAs)/AlGaAs QDs, which were obtained by infilling nanoholes created by in-situ droplet etching of AlGaAs layers (see Fig. \ref{fig:fig1} (a) and methods)\cite{Huo:APL2013,APL1.3653804}. Atomic force microscopy of a representative nanohole (see Fig. \ref{fig:fig1} (c)) reveals a highly symmetric shape. This shape, which determines the QD confining potential, is of fundamental importance for the generation of entangled photons. Fig. \ref{fig:fig1} (d) shows a typical microphotoluminescence ($\mu-$PL) spectrum of a single GaAs QD under non-resonant excitation. It consists of an isolated X line at high energy and a bunch of neutral and charged multiexcitonic transitions at lower energy (roughly 4 meV  below the X). While the detailed discussion of the origin of these states is beyond the scope of this work, we would like to point out that the XX transition is almost always overwhelmed by them and cannot be singled out easily under non-resonant pumping, even at the highest excitation power. This observation is qualitatively ascribed to the large number of confined states in the employed QDs, which lead to many possible charge configurations competing with the XX at high excitation power. In order to make the XX line visible and to gain coherent control over this state, we use a two-photon resonant excitation (TPE) scheme. More specifically, we shape the excitation laser into 9 ps pulses (see methods) and we tune its energy to half of the energy between the XX and ground state (0) (see inset Fig. \ref{fig:fig1} (e)).

\begin{figure}[htbp]
	\centering
		\includegraphics[width=180mm]{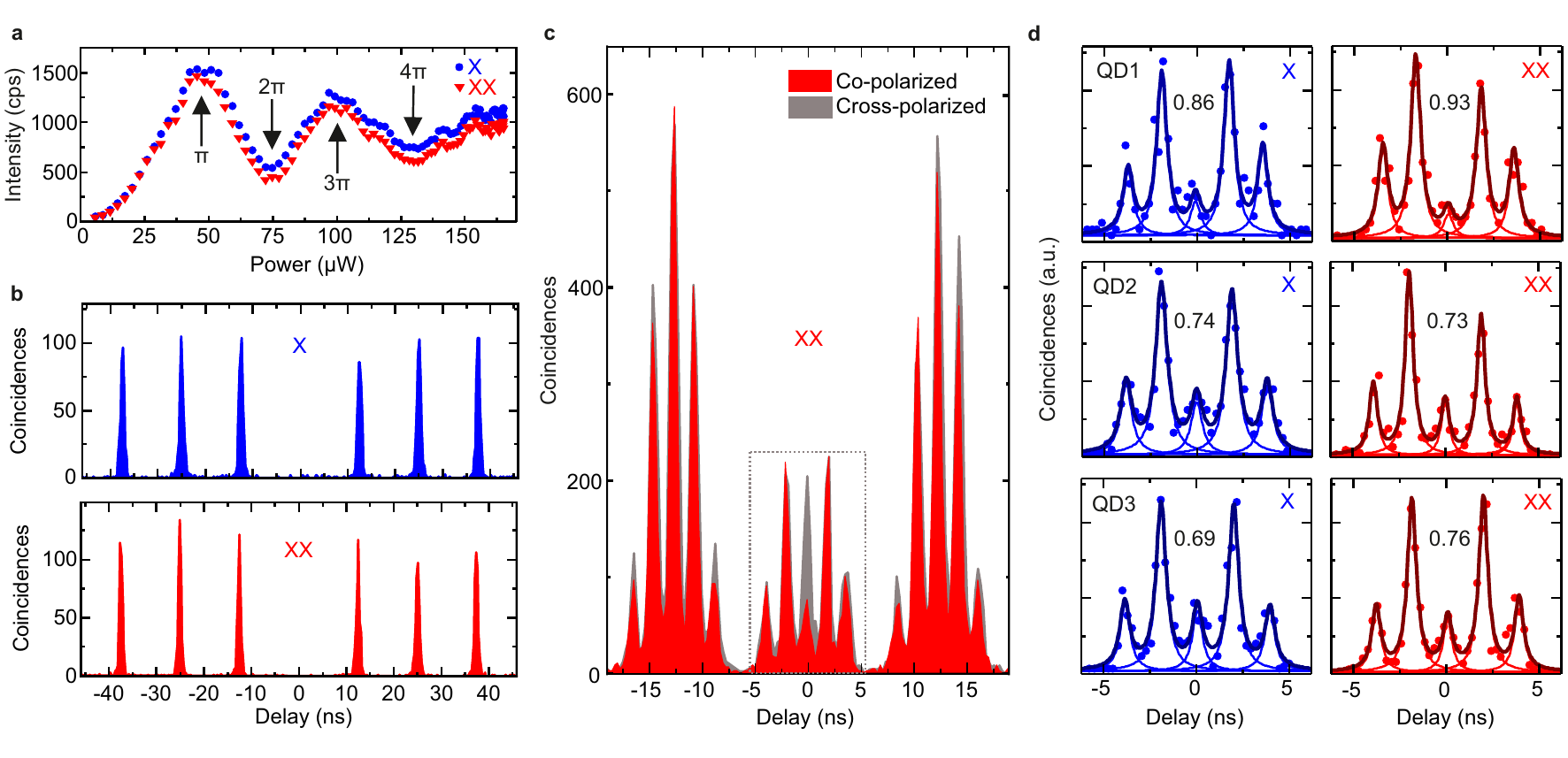}
	\caption{\textbf{Single photon purity and two-photon interference.} (a) Power-dependent measurements under resonant excitation of a representative QD. The X (blue circles) as well as the XX (red circles) show well defined Rabi oscillations, up to 4$\pi$. (b) Auto-correlation measurements of X and XX photons for a representative QD. The measurements show a strong anti-bunching for both lines. (c) Two-photon interference in co- (red) and cross-polarized (grey) XX photons. (d) Same as in (c) for three different QDs around zero time delay (see gray box in c). The data for both the X (left panel) and XX (right panel) co-polarized photons are reported. The solid thin lines are Lorentzian fits of the peaks. The thick solid lines show their sum. The values of the visibility are also reported in each panel. 
	}
	\label{fig:fig2}
\end{figure}

The effect of the TPE in the emission spectrum is shown in Fig. \ref{fig:fig1} (e). The multiexcitonic states disappear and the XX line becomes visible with an intensity equal to that of the X. Two additional transitions appear (C1 and C2), which are most probably related to a charged exciton and biexciton. These states, whose detailed investigation is left to future studies, are present in all the QDs we measured. To verify whether the QD is driven really resonantly, we measured the power dependence of the intensity of X and XX (see Fig. \ref{fig:fig2} (a)). Both transistions show well-defined Rabi oscillations up to pulse areas of 4$\pi$, thus readily confirming that our excitation scheme allows for coherent control of the XX state. The damping of the Rabi oscillations most probably arises from phonon-induced dephasing, as already observed in the literature\cite{Michler:NaturePhoton2014}. For all subsequent measurements the power has been set to the $\pi$-pulse condition.

We now characterize the properties of the photons emitted by our QDs in terms of their purity and indistinguishability. In Fig. \ref{fig:fig2} (b) we report auto-correlation measurements performed on the X and XX lines, which allows us to estimate the value of the second-order correlation function at zero time delay, being g$^{(2)}(0)=0.007\pm0.004$ and g$^{(2)}(0)=0.002\pm0.002$ for X and XX, respectively. This result clearly indicates an extremely high purity of our source.

Next, we study the photon indistinguishability of consecutive photons emitted by our QDs by performing an Hong-Ou-Mandel type two-photon interference experiment\cite{Ou:PhysRev1988}. The QDs are excited every 12.5 ns by two $\pi$-pulses separated 2 ns and emit ideally two XX-X photon pairs per cycle. The photons are then guided to an unbalanced Mach-Zehnder (MZ) interferometer equipped with 2 ns delay and are let interfere at the beam splitter (BS) in co- and cross-polarized configuration. 

An example of the measurements for the XX are shown in Fig. \ref{fig:fig2} (c). For cross-polarized photons, the three central peaks around 0 time-delay show the same amplitude, as expected for distinguishable photons. Contrarily, the strong suppression of the central peak for co-polarized photons clearly indicates the occurrence of two-photon interference and pinpoints to a large indistinghuishability. A similar behavior is observed for X photons, as shown in the of Fig. \ref{fig:fig2} (d). By Lorentzian fitting (see Supplementary Note 2) and by taking into account the imperfections of the BS\cite{Yoshihisa:Nature2002,Senellart:NatPhoton2016,Michler:NaturePhoton2014}, we estimate visibilities of two-photon interference of $0.86\pm 0.09$ and $0.93\pm 0.07$ for X and XX photons, respectively (details on the calculations see Supplementary Note 2). These values are the highest reported so far for QDs driven under two-photon excitation and are very close to the values achieved for near-optimal single-photon sources\cite{Senellart:NatPhoton2016,PhysRevLett.116.020401}. While the slightly lower value of visibility of the X can be explained by the time jitter introduced by the XX radiative decay, we emphasize that this is not a general feature of our QDs, as discussed below. 

We performed the same measurements in several additional QDs and, for some of them, we report the relevant results in Fig. \ref{fig:fig2} (d). Large visibility (exceeding 70$\%$) of two-photon interference has been found in almost all the QDs we measured, for both X and XX photons. The slightly different values observed can be explained by the different charge-noise configuration in each specific QD, as also reported in the literature\cite{Senellart:NatPhoton2016}. 

\begin{figure}[htbp]
	\centering
		\includegraphics[width=180mm]{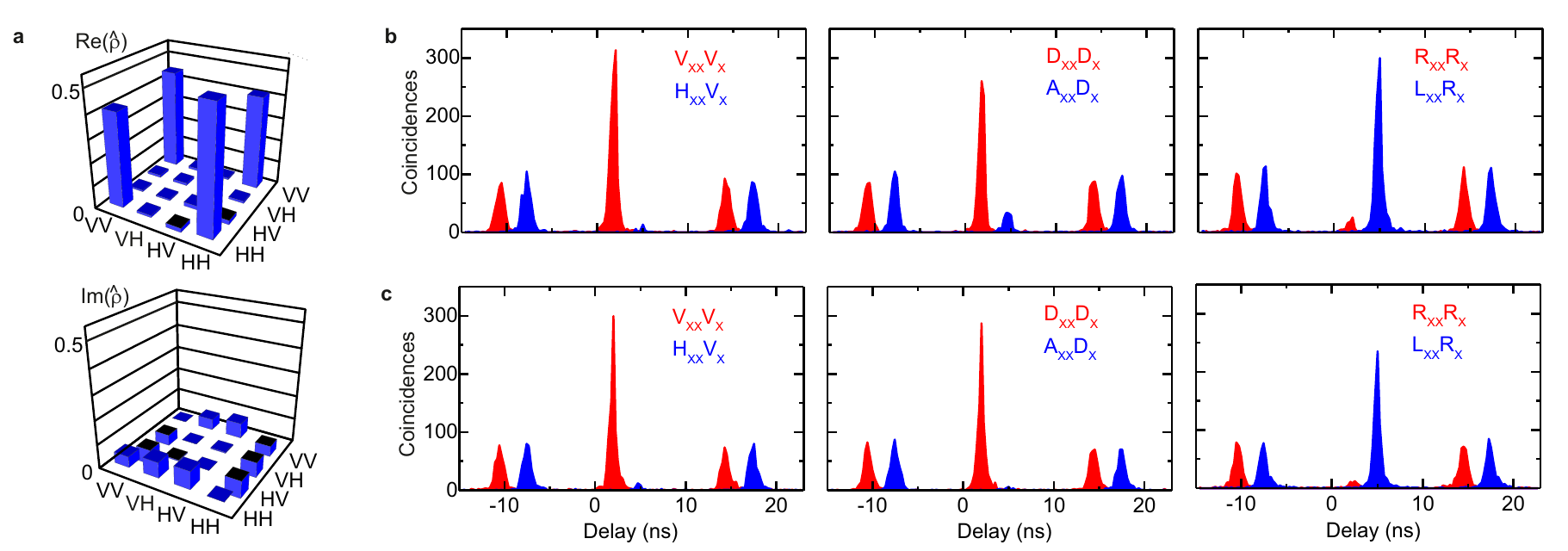}
	\caption{\textbf{Entangled photons from GaAs QDs.} (a) Real and imaginary part of the two-photon density matrix reconstructed via quantum state tomography on QD2. The matrix has been calculated out of XX-X cross-correlation measurements in 16 polarization settings and using a maximum likelihood method. (b) XX-X cross-correlation measurements under resonant two-photon excitation for QD2 and (c) QD3. V$_{XX,X}$ H$_{XX,X}$,D$_{XX,X}$ A$_{XX,X}$ and R$_{XX,X}$ L$_{XX,X}$ indicate vertical (horizontal), diagonal (anti-diagonal) and circularly-right (circularly-left) polarized photons. The graphs for co- (red) and cross- (blue) polarized photons are temporally shifted by 3 ns for clarity.
}
	\label{fig:fig3}
\end{figure}

Finally, we complete our study demonstrating that our QDs emit photon pairs with unprecedented degree of entanglement. Here we take advantage of two features of our QDs: (i) their extremely high structural symmetry (see Fig. \ref{fig:fig1} (a)), which results in an extremely small average fine-structure splitting (FSS) between the two X states\cite{Huo:APL2013} and (ii) the short radiative decay time of X ($T_{1}$ around 250 ps, see the Supplementary Note 1). It is indeed well known that only if the FSS is smaller or comparable to the radiative linewidth of the transitions, a large degree of entanglement can be measured in a time integrated experiment\cite{Stevenson:PRL2008}. The FSS of QD1,2,3 are, respectively 6.5$\pm0.5$ $\mu$eV, 1.3$\pm0.5$ $\mu$eV and 1.2$\pm0.5$ $\mu$eV. Therefore, we do not expect a high degree of entanglement for QD1 (see Suplementary Note 5) and only QD2 and QD3 are retained for the following study. 

We start out with QD2 and perform a full quantum-state tomography\cite{White:PRA2001} to reconstruct the two photon density matrix (see Fig. \ref{fig:fig3} (a)) with the aid of a maximal likelihood method\cite{White:PRA2001}. In order to quantify the degree of entanglement, we extract the concurrence C from the density matrix and find $C=0.84\pm 0.05$, the highest value ever reported for QD photon sources. Closer inspection of the density matrix reveals that the eigenstate associated to the largest eigenvalue (0.92$\pm 0.03$) is not exactly the expected Bell state $\ket{\psi^+}=1/\sqrt{2}\ket{H_{XX}}\ket{H_{X}}+\ket{V_{XX}}\ket{V_{X}}$\cite{Stevenson:PRL2008}, but rather $\ket{\psi_{r}^+}\approx 1/\sqrt{2}\ket{H_{XX}}\ket{H_{X}}+e^{i\alpha\pi}\ket{V_{XX}}\ket{V_{X}}$, with $\alpha=0.12$. The phase is most probably induced by the measurement setup and in particular by the presence of background photons of the resonant laser that could not be suppressed completely during quantum state tomography (see Supplementary Note 3).

In order to demonstrate that the high entanglement degree is not a feature of one specific QD, we extend the study to QD3. In this case we took special care in rejecting the resonant laser and we extracted the fidelity to expected Bell state via

\begin{equation}
      f=\frac{1+C_{linear}+C_{diagonal}-C_{circular}}{4},
			\label{eq:fidform}
\end{equation} 

where C's are the correlation visibilities in the linear, diagonal, and circular bases (see Supplementary Note 4). In fact, for QD2 (see Fig. \ref{fig:fig3} (b)) we confirmed that the value of the fidelity obtained with this formula (0.88$\pm$0.01) is very close to the one extracted via full reconstruction of the density matrix (0.92$\pm 0.03$) and, since $\alpha$ is small, to the largest eigenvalue (0.92$\pm0.03$). We have therefore used the 6 measurements reported in Fig. \ref{fig:fig3} (c) to extract the fidelity from Eq.\ref{eq:fidform}, which is found to be $0.94 \pm 0.01$. This value -  which is higher than the one measured for QD2 due to improved rejection of the laser (see Supplementary Note 3 for comparison of the $g^{(2)}(0)$ values) -  is very close to the maximum achievable (see Supplementary Note 5). It is worth mentioning that we obtain this results without any temporal/spectral filtering of the emitted photons and without the aid of the Purcell effect\cite{PhysRevLett.116.020401}, which can in principle be used to shorten even further the X radiative decay time, and to improve further the level of entanglement and indistinghuishability to the maximum. The high degree of entanglement allows for violation of Bell's inequality. Using the following expressions:

\begin{equation}
\begin{aligned} 
    S_{RC}=\sqrt{2}(C_{linear}-C_{circular})\leq 2 \\
		S_{DC}=\sqrt{2}(C_{diagonal}-C_{circular})\leq 2 \\
		S_{RD}=\sqrt{2}(C_{linear}+C_{diagonal})\leq 2
\end{aligned}  
\end{equation} 

we obtain $S_{RC}=2.52 \pm 0.02$, $S_{DC}=2.59 \pm 0.01$ and $S_{RD}=2.64 \pm 0.01$. The latter parameter shows violation of Bell's inequality by more than 60 standard deviations, a result that we achieve without any temporal or spectral filtering.

The measurements of Fig. \ref{fig:fig3} (c) readily suggest the origin of this unprecedented level of entanglement. In fact, an almost perfect antibunching is observed for cross-linear, cross-diagonal and co-circular base. This result clearly indicates a strong suppression of X spin-flip processes\cite{PhysRevLett.99.266802} and, most importantly, of recapture, a result which is most probably a consequence of the resonant excitation scheme used. Moreover, the fact that the correlation visibility is lower in the circular than in the linear$/$diagonal base suggests that the X dephasing induced by the hyperfine interaction is quite limited in these QDs\cite{Burk:arxiv2015}. In fact, a theoretical model\cite{PhysRevLett.99.266802} which takes into account the effect of the Overhauser field on dephasing of the intermediate X states (see Supplementary Note 5) shows that this is exactly the reason why our GaAs QDs show such a high level of entanglement. The same model can be also used to highlight that the small deviation from unity is mainly ascribable to the non-zero values of the FSS. More specifically, we estimate that in GaAs QDs with suppressed FSS - reachable, e.g., via three-axial strain engineering\cite{Trotta:2016NatCom} - the fidelity can reach values as high as 0.99, roughly 10$\%$ higher than what can be theoretically achieved with InGaAs QDs with FSS=0 with the same X decay time (see Supplementary Note 5).

Our results show that the GaAs/AlGaAs material system, which has led to breakthrough discoveries in condensed matter physics and opto-electronics (such as the fractional quantum Hall effect\cite{WILLETT1988257}) has also a great potential for the implementation of as entanglement resources for long distance quantum communication, and in particular for quantum repeaters\cite{Kimble:Nat2008}. Beside the near-optimal level of entanglement and indistinguishability of the emitted photons - fundamental requirements to teleport entanglement over distant parties - our QDs emit in the spectral range of absorption resonances of Rb atoms, a system which has proven its potential as long-lived and efficient quantum memory\cite{Kupchak:SciRep2015}. To really use these QDs in real-life applications, the next step is to integrate them in photonic cavities for boosting the flux of the XX and X photons and to gain control over the energy of the emitted photons using external perturbations\cite{Trotta:2016NatCom}. These are challenging tasks, but the strive to overcome them is worth the efforts, as the implementation of of the ideal source of entangled photons is expected to be revolutionary.

\section*{Methods}

\subsection*{Sample growth}

The sample is grown by solid state molecular beam epitaxy (MBE). A GaAs (001) substrate is overgrown with a 270 nm buffer layer, on which 3 pairs of alternating $Ga_{0.8}Al_{0.2}As/AlAs$ (with a thickness of 56 nm and 65 nm) are deposited. On top of this layer structure, which forms a low reflectivity back mirror of a planar cavity, a 120 nm thick layer of $Ga_{0.6}Al_{0.4}As$ is grown. This layer is hosting the nanoholes, which are fabricated at 600 $^\circ\text{C}$ using Aluminum droplet etching\cite{Huo:APL2013}. The QDs are then obtained by filling the nanoholes with $2.5$ nm of GaAs, subsequently capped with 120 nm of $Ga_{0.6}Al_{0.4}As$ acting as top barrier. Finally, two pairs of $Al_{0.2}Ga_{0.8}As/AlAs$ layers were grown to complete the cavity. Due to the limited number of pairs used for the distributed Bragg reflector (DBR) mirrors, we do not expect any appreciable Purcell effect but only a slight enhancement of the light extraction efficiency compared to a sample without DBRs. 

\subsection*{Measurement Setup}

All the measurements are performed at a sample temperature of 5 K in a helium-flow cryostat. 
For non-resonant excitation, a 488 nm continuous wave (cw) laser is used. The resonant two-photon excitation is performed with a titanium sapphire (TiSa) femtosecond laser featuring bandwidth of 100 fs and a repetition rate of 80 MHz. The TiSa is shaped by a 4f pulse-shaper setup into 9 ps pulses. The excitation lasers are focused via an objective with a numerical aperture of 0.42 onto the sample, on which a hyperspherical solid immersion lens, made of zirconia, is placed. The same objective is used for the collection of the PL signal.

To reduce the back scattered laser light in resonant excitation, tunable notch filters with a bandwidth of 0.4 nm are placed in the beam path. Further stray light reduction is achieved by coupling the signal into polarization maintaining (PM) single mode fibers placed after a non-polarizing beam splitter.
Depending on the QD, the addition of the CW laser at extremely low (not measurable) power to the resonant laser increases the X and XX count rate. This non-resonant laser - which does not give rise to background photons -  most probably fills traps in the QD surrounding. 

After the fibers, light is guided to and spectrally filtered by one or two independent spectrometers and detected by a CCD camera or - during photon correlation spectroscopy - by avalanche photodiodes (APD) connected to the correlation electronics. The temporal resolution of these detectors is about 500 ps.

The measurements of the FSS are carried out with a rotating $\lambda/2$ wave plate in front of a fixed polarizer at the entrance of a double spectrometer. Using Lorentzian fitting of both the X and XX lines, the FSS can be measured with sub-$\mu$eV resolution. For quantum state tomography, polarizers and properly oriented  $\lambda/2$ and $\lambda/4$ wave plates are inserted in the X and XX path, before the two fiber inputs.

For the two-photon interference experiment, the femtosecond laser pulses are sent into an unbalanced MZ with a 2 ns delay to create two pulses every 12.5 ns. Both laser pulses are shaped with the 4f-pulseshaper described above. A polarization maintaining fiber BS is used for two-photon interference of QD photons. The measurements for co-linear and cross-linear configuration are obtained by placing polarizers before the input of the fiber-BS. The latter is characterized with the same pulse shaped laser used to excited the QDs and shows a mode overlap $(1-\epsilon)=0.96\pm 0.01$, while the reflection and transmission coefficient are found to be $0.52\pm 0.005$ and $0.48\pm 0.005$, respectively.      

\newpage

\bibliography{main4}

\section*{Acknowledgements}

This work was financially supported by the ERC Starting Grant No. 679183 (SPQRel) and European Union Seventh Framework Programme (FP7/2007-2013) under grant agreement no. 601126 (HANAS).

We acknowledge Dr. Javier Martín-Sánchez and Dr. Klaus Jöns for fruitful discussions and for help at the very early state of the project. 

\section*{Author contributions statement}

D.H. and M.R. performed the measurements with the help from R.T.. D.H. made the data analysis with the help from M.R.,J.S.W. and R.T.. Y.H. grew the sample with support of A.R. and O.G.S.. H.H. and M.R. characterized the sample. D.H. and R.T. wrote the manuscript with the help from all the authors. R.T. and A.R. conceived the project. R.T. designed the experiments and supervised the project.

\section*{Competing financial interests}
The authors declare no competing financial interests.

\end{document}